\documentclass[doublecol]{epl2} 

\usepackage{amssymb}
\usepackage{mathtools}
\usepackage{amsmath}
\usepackage[utf8]{inputenc}
\usepackage[english]{babel}
\title{Systems of random linear equations and the phase transition in MacArthur's resource-competition model}
\shorttitle{On the phase transition in MacArthur's resource-competition model} 

\author{Stefan Landmann\inst{1}\hspace{-0.2cm}\footnote{E-Mail: \email{stefan.landmann@uni-oldenburg.de}} \and Andreas Engel\inst{1}}
\shortauthor{Stefan Landmann \etal}

\institute{                    
	\inst{1} Carl von Ossietzky Universit\"at, Institut f\"ur Physik - D-26111 Oldenburg 
	Germany\\
}

\abstract{
	Complex ecosystems generally consist of a large number of different species utilizing a large number of different resources. Several of their features cannot be captured by models comprising just a few species and resources. Recently, Tikhonov and Monasson have shown that a high-dimensional version of MacArthur's resource competition model exhibits a phase transition from a 'vulnerable' to a 'shielded' phase in which the species collectively protect themselves against an inhomogeneous resource influx from the outside. Here we point out that this transition is more general and may be traced back to the existence of non-negative solutions to large systems of random linear equations. Employing Farkas' Lemma we map this problem to the properties of a fractional volume in high dimensions which we determine using methods from the statistical mechanics of disordered systems. 
}

\begin{document}
	
	\maketitle

	\section{Introduction}
	Ecosystems -- from rainforests to the human gut -- can harbor a surprisingly large number of different species \cite{hutchinson1959homage,ghazoul2010tropical,lozupone2012diversity}. These species in general compete for a limited number of resources and possibly prey on each other. Inspired by this observation researchers from various fields examine the role biodiversity plays in complex ecosystems and how their community structure is shaped. However, mathematical studies of model ecosystems were mostly performed for systems with only a few species and resources \cite{macarthur_species_1970,tilman1982resource,huisman1999biodiversity}. Results obtained for small settings do not straightforwardly generalize to large systems characterized by collective phenomena and emergent properties \cite{levins1966strategy,anderson1972mor}. Starting with the pioneering work of May \cite{may1972will} such phenomena are increasingly addressed by studying large models with random parameters. This is a sensible approach if self-averaging properties of the system may be identified that only depend on the features of the underlying distributions and not on the individual realization of the randomness. Methods from the statistical mechanics of disordered systems then provide useful tools for a quantitative characterization of typical properties of the system \cite{TM17,tikhonov2017innovation,biroli2017marginally,advani2017environmental}. 
	
	Along these lines, Tikhonov and Monasson \cite{TM17,tikhonov2016community} recently investigated a high-dimensional version of MacArthur's consumer-resource model \cite{macarthur_species_1970}. In this model, different species compete for a number of resources which are supplied by fixed influxes from the environment. An increasing population size of a species leads to a lower resource availability and consequentially to a reduced growth rate, creating a negative feedback loop. Even though the interactions between the species are purely competitive Tikhonov and Monasson found a transition into a collective phase at large potential diversity, i.e., when the number of available species in the regional pool sufficiently exceeds the number of resources. In this phase, all resources are equally well available  and the number of surviving species is equal to the number of resources, saturating the upper bound set by the competitive exclusion principle \cite{armstrong1980competitive}. Moreover, by performing a stability analysis Altieri and Franz \cite{altieri2018constraint} revealed that the collective phase exhibits marginally stable behavior.
	
	In \cite{TM17} the phase transition was examined by constructing a Lyapunov-function for the population dynamics and a subsequent replica calculation characterizing its extremum. In the present letter we show that the transition is more general and may be derived without reference to the actual dynamics of the model. We first point out a connection between stationary states of MacArthur's model and the space of non-negative solutions to large systems of random linear equations. By the Farkas lemma this solution space is related to a random fractional volume in high dimensions the typical value of which we then determine.

	\section{The model}
	
\begin{figure}
		\centering
		\includegraphics[width=\linewidth]{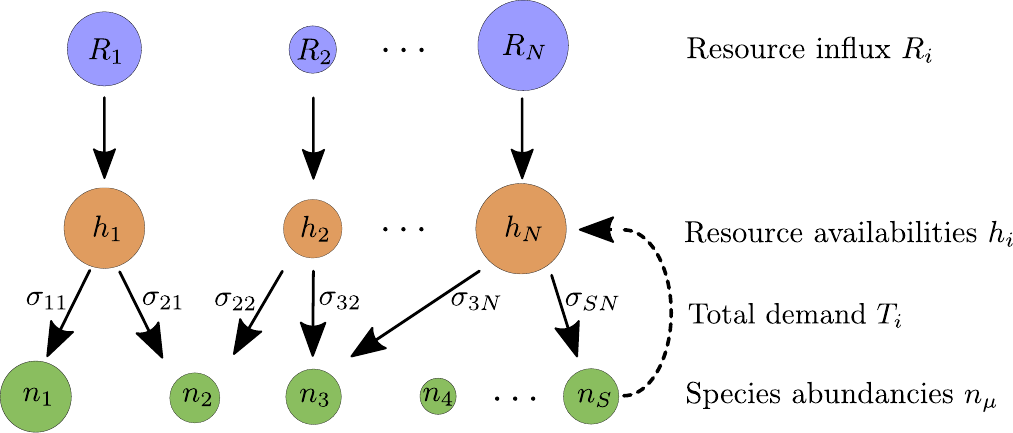}
		\caption{
			Schematic of the resource-competition model. The fixed external resource influxes $R_i$ get modified to the resource availabilites $h_i$ depending on the overall demand $T_i$ of resource $i$. If many species with high abundancies use the same resource, its availability will decrease; if only a few species with low abundancies can use a resource, its availability will be high. Species are characterized by their metabolic strategies  $\sigma_{\mu i}$ that specify which resource a species may consume.  Increasing the abundancy of one species will always imply a slower growth of the other species, the interaction is purely competitive.}
		\label{fig:schematic of model}
	\end{figure}

	The version of MacArthur's resource-competition model considered in \cite{TM17} consists of $S$ species with abundancies $n_\mu\geq 0,\, \mu=1,\dots, S$ which can utilize $N$ resources with availability $h_i,\, i=1,\dots, N$. 
	The resources are supplied from the outside by fixed influxes $R_i$ that, depending on the overall demand $T_i$, are reduced to the resource availabilites $h_i$, cf. Figure~\ref{fig:schematic of model}. Since we are interested in high-dimensional situations we consider the combined limit $N\to \infty,\, S\to \infty$ with the ratio, $\alpha:=S/N$ staying constant. As one of the main parameters of the model $\alpha$ specifies the potential diversity of the ecosystem. 
	
	Each species $\mu$ is characterized by a metabolic strategy vector $\boldsymbol{\sigma}_\mu \in \mathbb{R}^N $ specifying which resources it can employ. Its entries $\sigma_{\mu i}$ are one if it may consume resource $i$ and zero otherwise. Moreover, each species needs a minimal resource supply $\chi_\mu$ in order to survive: if the resource intake is smaller than $\chi_\mu$, its population shrinks; otherwise it grows. The system dynamics is described by a Malthusian law
	
	\begin{equation}
	\frac{dn_\mu}{dt}=n_\mu \left[\sum_i \sigma_{\mu i} h_i - \chi_\mu \right].
	\label{eq:Dynamics}
	\end{equation}
	
	In order to allow for a fair competition between specialists and universalists the threshold $\chi_\mu$ is chosen to increase with the number of utilizable resources \cite{remark}, 
	
	\begin{equation}\label{eq:defchi}
	\chi_\mu=\sum_i \sigma_{\mu i}.
	\end{equation} 
	
	The resource availabilities $h_i$ derive from the resource influxes $R_i$ and are decreasing functions of the total demand $T_i=\sum_\mu \sigma_{\mu i} n_\mu$ of resource $i$. Different models of resource supply differ in the form of these depletion functions $h_i(T_i)$. We require that if influx and total demand balance for every resource the dynamics \eqref{eq:Dynamics} should be in a stationary state which according to \eqref{eq:defchi} corresponds to $h_i=1$ for all $i$. The dependence of the resource availabilities on the total demand is hence given by  
	\begin{equation}
	h_i(T_i)=1-f_i(T_i),
	\label{eq:resource availability}
	\end{equation}
	where the functions $f_i$ are monotonically increasing functions of their argument and satisfy  $f_i(R_i)=0$. We do not need any further specification of the depletion functions in our analysis. 
	
	In line with previous investigations \cite{may1972will,tokita2015analytical,TM17} it is assumed that the model parameters $\sigma_{\mu i}$ and $R_i$ are independent random variables. More specifically, $\sigma_{\mu i}$ is taken to be one with probability $p$ and zero with probability $1-p$. Small values of $p$ therefore describe populations with many specialists whereas large $p$ favours universalists. The $R_i$ are taken to be of the form $R_i=1+\delta R_i$ where the fluctuations $\delta R_i$ are Gaussian random variables with zero mean and variance $r^2/N$. The scaled variance $r^2$ remains $O(1)$ for $N\to\infty$ and characterizes as a second central parameter of the model the heterogeneity of resource influxes to the system.
	
	For linearized depletion functions \eqref{eq:resource availability} it was shown in \cite{TM17} that the system possesses two different phases. For small potential diversity, $\alpha\leq\alpha_c$, the system is in the vulnerable or V-phase. Here the inhomogeneity of resource influxes $R_i$ penetrates down to the level of the resource availabilities $h_i$ and the number of surviving species is less than $N$. In contrast, in the shielded or S-phase at large potential diversity $\alpha$, all resource availabilities $h_i$ are equal to one despite the differences in the external influxes $R_i$.  In this phase the species form a kind of collective field and shield each other from the external inhomogeneities. At the same time the number of surviving species attains its maximum value $N$. 
	
	In their calculation Tikhonov and Monasson construct a convex Lyapunov function $F( \mathbf{n})$ which is bounded from above and increases on every trajectory. They show that the stationary states of the system lie on the boundary of the so-called unsustainable region in the space of resource availabilities 
		\begin{equation}
		U=\bigcap^S_{\mu=1} \{ \mathbf{h} \, | \, \boldsymbol{\sigma}_\mu \cdot \mathbf{h} < \chi_\mu \}.
		\end{equation}
		A partition function of the system is then defined by 
		\begin{equation}
		Z=\int_{U} \prod_i dh_i e^{\beta \tilde{F}(\mathbf{h})},
		\end{equation}	
		where the integration is performed over the unsustainable region and $\tilde{F}(\mathbf{h})$ is the Legendre transform of $F(\mathbf{n})$. In the limits $\beta,N \rightarrow \infty $ the expectation value $\frac{1}{\beta N}\left< \log Z \right>$ is determined by application of the replica trick, thereby characterizing the stationary states and determining the critical line of the phase transition. In the following we present a different approach to the transition which does not make use of the Lyapunov function.

	\section{Relation to systems of linear equations}
	
	\begin{figure}[t!]
		\centering
		\includegraphics[width=\linewidth]{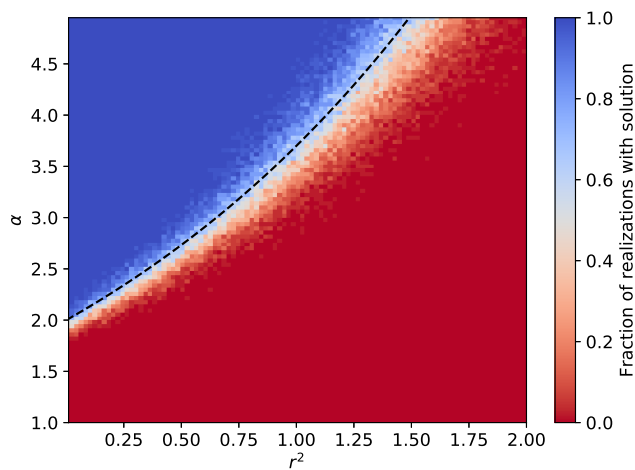}
		\caption{Average fraction of realizations of Eq.~(\ref{eq:linear equation}) which possess a non-negative  solution $\mathbf{n}$. The solution space clearly separates into a part with typically no such solution (lower right) and a phase in which such a solution typically exists. The dashed line is the critical line determined in \cite{TM17} and also given by Eq.~\eqref{eq:Critical Line} below. The system size is $N=300$, $p=0.5$, and each data point was averaged over 50 realizations.}
		\label{fig:Fraction of solutions}
	\end{figure}
	
	\begin{figure}[]
		\vspace{-0.4cm}
		\centering
		\includegraphics[width=\linewidth]{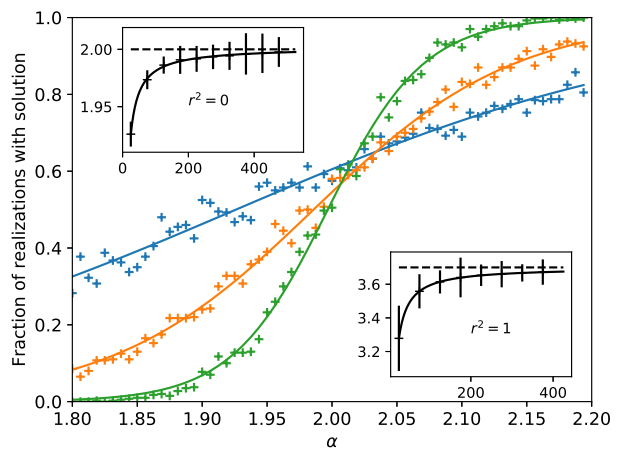}
		\vspace{0.0cm}
		\caption{Finite size analysis of the critical value $\alpha_c$ at which the phase transition occurs. The figure shows the fraction of realizations of Eq.~(\ref{eq:linear equation}) which possess a non-negative solution for $r^2=0$ and $N=25, \, 125$ and  $525$, respectively, ordered by increasing steepness. Each data point was averaged over 400 realizations. The insets show the finite-size approximations to $\alpha_c$ plotted as function of $N$ for $r^2=0$ (upper left) and $r^2=1$ (lower right). The dashed lines are the analytical results from Eq.~\eqref{eq:Critical Line}. $p=0.5$ for all cases shown here. For further details see the supplementary material.}
		\label{fig:Finite Size}
	\end{figure}

	The phase transition observed in the resource-competition model can be related to a problem in linear algebra. Let us denote by $\Hat{\sigma}$ the matrix $\sigma_{\mu i}$ of metabolic strategies, by $\mathbf{n}\in\mathbb{R}^{\alpha N}$ the vector of species abundancies, and by $\mathbf{R}\in\mathbb{R}^N$ the vector of resource influxes. The central point is that the S-phase is characterized by $h_i=1$ for all resources $i$. By \eqref{eq:resource availability} this implies $T_i=R_i$ for all $i$ and using the definition of $T_i$ it translates to 
	\begin{equation}\label{eq:linear equation}
	 \Hat{\sigma}^T \mathbf{n}=\mathbf{R}.
	\end{equation} 
	We therefore expect that the system is in the S-phase if these inhomogeneous linear equations possess a {\em non-negative} solution $\mathbf{n}, n_\mu\geq 0$ and that it is in the V-phase if no such solution exists. Figure~\ref{fig:Fraction of solutions} tests this assumption on the basis of numerical simulations. Shown is the fraction of random realizations of Eq.~(\ref{eq:linear equation}) for which a non-negative solution was found by application of a least squares solver \cite{usedsolver}. It is clearly seen that the solution space of Eq.~(\ref{eq:linear equation}) is separated into a phase in which typically no solution exists and a phase in which a solution can always be found. The dashed line marks the phase transition derived in \cite{TM17}. Similar behavior is found for other values of $p$.
	
	The observed transition becomes sharper when the systems gets larger as shown by the finite size analysis of  Fig.~\ref{fig:Finite Size}. As can be seen the steepness of the transition increases with increasing system size $N$ and the extrapolated values of $\alpha_c$ converge to the analytical result given by Eq.~\eqref{eq:Critical Line}.

	\section{Analytical determination of the critical line}
	The question whether the linear system \eqref{eq:linear equation} for large $N$ typically possesses a non-negative solution $n_\mu$ can be analyzed analytically. To this end we first employ Farkas' Lemma \cite{farkas1902theorie} that stipulates that for given $\Hat{\sigma}$ and $\mathbf{R}$ either \eqref{eq:linear equation} has a non-negative solution or there is a vector $\mathbf{y}\in \mathbb{R}^N$ such that 
	\begin{equation}\label{Eq: Farkas Lemma} 
	\Hat\sigma \mathbf{y} \geq 0\qquad\mathrm{and}\qquad \mathbf{R}\cdot\mathbf{y}<0. 
	\end{equation} 
	The intuitive meaning of this theorem is simple: the linear combinations of the row vectors $\boldsymbol{\sigma}_\mu$ of $\Hat{\sigma}$ with non-negative coefficients form what is called the non-negative  cone of these vectors. If $\mathbf{R}$ lies within this cone, there is a non-negative solution to Eq.~\eqref{eq:linear equation}; if not, there must be a hyperplane (with normal vector $\mathbf{y}$) separating the cone from $\mathbf{R}$. 
	
	The dual problem defined by \eqref{Eq: Farkas Lemma} is rather similar to the storage problem in the theory of feedforward neural networks \cite{engel2001statistical} and can be addressed by similar means. We define the fractional volume of vectors $\mathbf{y}$ that fulfil Eq.~(\ref{Eq: Farkas Lemma})
	\begin{align}
		\Omega(\Hat{\sigma},\mathbf{R}):=\frac{\int^\infty_{-\infty} \prod_i dy_i\, \delta \left(\sum_i y_i^2-N\right) \mathbf{1}(\mathbf{y};\hat{\sigma},\mathbf{R}) }{\int^\infty_{-\infty} \prod_i dy_i\, \delta(\sum_i y_i^2-N)},
	\label{Equ: Volume of solutions}
	\end{align}
	with the indicator function
	\begin{align}
	\mathbf{1}(\mathbf{y};\hat{\sigma},\mathbf{R}):= \prod^{\alpha N}_{\mu=1} \Theta \left( \frac{1}{\sqrt{N}} \sum_i \sigma_{\mu i} y_i \right) \hspace{-0.1cm}\Theta \hspace{-0.1cm} \left(-\frac{1}{\sqrt{N}}\sum_i  R_i y_i \right).
	\end{align}
	The spherical constraint $\sum_i y_i^2=N$ is introduced to lift the trivial degeneracy of solutions $\mathbf{y}\to\lambda\mathbf{y}$ for any positive $\lambda$. If $\Omega(\Hat{\sigma},\mathbf{R})$ is zero there are no solutions to \eqref{Eq: Farkas Lemma} and correspondingly there is a non-negative solution to \eqref{eq:linear equation}. Complementary, if $\Omega(\Hat{\sigma},\mathbf{R})$ is larger than zero, there are vectors $\mathbf{y}$ fulfilling \eqref{Eq: Farkas Lemma} and therefore no non-negative solution to \eqref{eq:linear equation} exists. The transition occurs when $\Omega(\Hat{\sigma},\mathbf{R})$ shrinks to zero. 
	
	Due to the product structure of $\Omega$ the entropy $\frac{1}{N}\log \Omega$ is expected to be self-averaging with respect to $\Hat{\sigma}$ and $\mathbf{R}$. We may hence characterize the typical situation in a large system by considering the average entropy  
	\begin{equation}
	S(\alpha,p,r^2):=\left< \log  \Omega \right>_{\hat{\sigma},\mathbf{R}}.
	\end{equation}
	With the help of the replica trick \cite{mezard1987spin} and using standard techniques \cite{engel2001statistical}
	this entropy may be expressed as a saddle-point integral over order parameters (see supplementary material)
	\begin{equation}
	m^a=\frac{1}{\sqrt{N}}\sum_i y^a_i\quad\mathrm{and}\quad 
	q^{ab}=\frac{1}{N}\sum_i y^a_i y^b_i. 
	\end{equation} 
	Within the replica-symmetric ansatz we find 
	\begin{align}\nonumber
	S(\alpha,p,r^2)=\mathrm{extr}&
	\left[\frac{1}{2}\log(1-q)+\frac{q}{2(1-q)}-\frac{\kappa^2(1-p)}{2pr^2(1-q)}\right.\\
	&\left.\qquad+\alpha\int Dt\, \log H \Big(\frac{\sqrt{q}\,t-\kappa}{\sqrt{1-q}}\Big)\right],
	\label{eq:Final entropy}
	\end{align}
	where the extremum is over $\kappa$ and $q$ and the abbreviations $Dt:=dt/\sqrt{2\pi}\,e^{-t^2/2},\; H(x):=\int_x^\infty Dt$ and $\kappa:=m\sqrt{p/(1-p)}$ were used. 
	\\
	At the transition the volume $\Omega$ shrinks to zero and the typical overlap $q$ between two different solutions $\mathbf{y}$ approaches one. Keeping only the most divergent terms of \eqref{eq:Final entropy} in this limit we find the following parametric representation of the critical line $\alpha_c(r^2)$ (see supplementary material):
	\begin{equation}
	r^2= \frac{1-p}{p}\frac{\kappa^2}{1-\alpha_c I(\kappa)}, \, \, \, \, \alpha_c H(\kappa)=1,
	\label{eq:Critical Line}
	\end{equation}
	where $I(\kappa):=\int_\kappa^\infty Dt\,(t-\kappa)^2$. This is the same result as found in \cite{TM17} exploiting the properties of the Lyapunov function. The expression for the entropy \eqref{eq:Final entropy} is rather similar to the one for the average entropy in the storage problem of a perceptron as obtained by Gardner \cite{gardner1988space}. In particular, for $r^2=0$ we find from \eqref{eq:Critical Line} $\kappa=0$  and therefore $\alpha_c=2$, the classical result for the storage capacity of the perceptron.
	
	\section{Conclusion}
	We have shown that the phase transition in a high-dimensional version of MacArthur's resource-competition model discovered recently by Tikhonov and Monasson is related to the existence of non-negative solutions of large random systems of linear equations. The starting point of our analysis is the observation that the 'shielded' phase in which the species collectively regulate the resource demand to make all resources equally available is also characterized by the maximally possible number of surviving species set by the competitive exclusion principle. In contrast, in the 'vulnerable' phase in which the species are susceptible to disturbances from the environment the number of surviving species remains below this margin. Since concentrations cannot be negative the difference between the two phases is related to the existence of non-negative solutions for species abundancies realizing appropriate resource availabilites. The transition depends on the ratio between the number of variables and the number of equations, the density of non-zero entries in the coefficient matrix and the variance of the inhomogeneity vector. The existence of non-negative solutions to underdetermined linear equations is an active field of research in its own. While prevalent techniques require sparseness of the solutions \cite{donoho2005sparse,wang2011unique} here we make -- in the limit where the number of unknowns and the number of equations tend to infinity -- also predictions for dense solutions. 
	
	Using Farkas' Lemma the question on the existence of non-negative solutions to linear systems can be mapped onto a dual problem involving a set of linear inequalities. Using methods from the statistical mechanics of disordered systems we have analytically analyzed the typical properties of this dual problem in the thermodynamic limit. The result is in perfect agreement with numerical simulations, reproduces the transition line found by Tikhonov and Monasson, and points out an interesting connection with the storage problem of the single-layer perceptron. This link may indicate a way to further improvements in the quantitative characterization of large random ecosystems. 
	
	\acknowledgments
	We would like to thank Remi Monasson, Katharina Janzen and Mattes Heerwagen for fruitful discussions. Financial support from the German Science Foundation DFG under grant EN 278/10-1 is gratefully acknowledged.
	\bibliography{bibliography.bib}{}

\begin{thebibliography}{10}
\expandafter\ifx\csname url\endcsname\relax\def\url#1{\texttt{#1}}\fi

\bibitem{hutchinson1959homage}
\Name{Hutchinson G.~E.} \REVIEW{The American Naturalist}{93}{1959}{145}.

\bibitem{ghazoul2010tropical}
\Name{Ghazoul J. \and Sheil D.} \Book{Tropical rain forest ecology, diversity,
  and conservation} (Oxford University Press) 2010.

\bibitem{lozupone2012diversity}
\Name{Lozupone C.~A., Stombaugh J.~I., Gordon J.~I., Jansson J.~K. \and Knight
  R.} \REVIEW{Nature}{489}{2012}{220}.

\bibitem{macarthur_species_1970}
\Name{MacArthur R.} \REVIEW{Theoretical Population Biology}{1}{1970}{1}.

\bibitem{tilman1982resource}
\Name{Tilman D.} \Book{Resource competition and community structure} (Princeton
  University Press) 1982.

\bibitem{huisman1999biodiversity}
\Name{Huisman J. \and Weissing F.~J.} \REVIEW{Nature}{402}{1999}{407}.

\bibitem{levins1966strategy}
\Name{Levins R.} \REVIEW{American scientist}{54}{1966}{421}.

\bibitem{anderson1972mor}
\Name{Anderson P.~W.} \REVIEW{Science}{177}{1972}{393}.

\bibitem{may1972will}
\Name{May R.~M.} \REVIEW{Nature}{238}{1972}{413}.

\bibitem{TM17}
\Name{Tikhonov M. \and Monasson R.} \REVIEW{Phys. Rev.
  Lett.}{118}{2017}{048103}.

\bibitem{tikhonov2017innovation}
\Name{Tikhonov M. \and Monasson R.} \REVIEW{Journal of Statistical
  Physics}{172}{2018}{74}.

\bibitem{biroli2017marginally}
\Name{Biroli G., Bunin G. \and Cammarota C.} \REVIEW{arXiv preprint
  arXiv:1710.03606}{}{2017}{}.

\bibitem{advani2017environmental}
\Name{Advani M., Bunin G. \and Mehta P.} \REVIEW{arXiv preprint
  arXiv:1707.03957}{}{2017}{}.

\bibitem{tikhonov2016community}
\Name{Tikhonov M.} \REVIEW{Elife}{5}{2016}{}.

\bibitem{armstrong1980competitive}
\Name{Armstrong R.~A. \and McGehee R.} \REVIEW{The American
  Naturalist}{115}{1980}{151}.

\bibitem{altieri2018constraint}
\Name{Altieri A. \and Franz S.} \REVIEW{arXiv preprint
  arXiv:1805.06412}{}{2018}{}.

\bibitem{remark}
In \cite{TM17} an additional random scatter $\epsilon x_\mu$ of these
  thresholds is introduced, where $x_\mu$ is a random Gaussian variable with
  $\left<x_\mu \right>=0$, $\left<x_\mu^2 \right>=1$ and $\epsilon$ is a small
  positive number. Since the sharp transition to the collective phase only
  occurs for vanishing scattering we consider the case $\epsilon \rightarrow
  0^+$. For further details confer the supplement.

\bibitem{tokita2015analytical}
\Name{Tokita K.} \REVIEW{Population Ecology}{57}{2015}{53}.

\bibitem{usedsolver}
We used the solver \textit{nnls} of the scipy.optimize package in Python.

\bibitem{farkas1902theorie}
\Name{Farkas J.} \REVIEW{Journal f{\"u}r die reine und angewandte
  Mathematik}{124}{1902}{1}.

\bibitem{engel2001statistical}
\Name{Engel A. \and Van~den Broeck C.} \Book{Statistical mechanics of learning}
  (Cambridge University Press) 2001.

\bibitem{mezard1987spin}
\Name{M{\'e}zard M., Parisi G. \and Virasoro M.} \Book{Spin glass theory and
  beyond} (World Scientific Publishing Company) 1987.

\bibitem{gardner1988space}
\Name{Gardner E.} \REVIEW{Journal of physics A: Mathematical and
  General}{21}{1988}{257}.

\bibitem{donoho2005sparse}
\Name{Donoho D.~L. \and Tanner J.} \REVIEW{Proceedings of the National Academy
  of Sciences of the U.S.A.}{102}{2005}{9446}.

\bibitem{wang2011unique}
\Name{Wang M., Xu W. \and Tang A.} \REVIEW{IEEE Transactions on Signal
  Processing}{59}{2011}{1007}.

\end{thebibliography}
	\bibliographystyle{eplbib}

\newpage
\onecolumn
\shorttitle{Supplementary material}
\setcounter{page}{1}
\setcounter{equation}{0}
\begin{center} 
\hspace{3 cm}\textbf{Supplementary material to}
\newline 
\\ 
\textit{Systems of random linear equations and the phase transition in MacArthur's resource-competition model}
\newline
\\ 
Stefan Landmann and Andreas Engel
\end{center}
\vspace{1cm}

	\section{The problem}
	The starting point of our considerations is the following question: Given an $\alpha N \times N$ random matrix $\hat{\sigma}$ and a random vector $\mathbf{R}\in\mathbb{R}^N$ when does the system of linear equations \begin{equation}
	\hat{\sigma}^T\mathbf{n}=\mathbf{R},
	\label{Equ: Linear Equation System}
	\end{equation}
	typically possess a solution $\mathbf{n}$ with all components non-negative, $n_\mu\geq 0,\, \mu=1,..,\alpha N$? The entries $\sigma_{\mu i}$ of $\hat{\sigma}$ are independently of each other one with probability $p$ and zero with probability $1-p$. The components of the vector $\mathbf{R}$ are of the form $R_i=1+\delta R_i$ with the $\delta R_i$'s being independent Gaussian variables with zero mean and variance $\frac{r^2}{N}$ where $r^2=O(1)$ when $N$ grows large.
	
	\section{Farkas' Lemma}
	Farkas' Lemma states that for a real matrix $\hat{\sigma}$ and a real vector $\mathbf{R}$ always one but only one of the following systems has a solution:
	\begin{align} \label{eq:n}
	& \, \,  \hat{\sigma}^T\mathbf{n}=\mathbf{R}, \,  \, \text{with} \, \,\mathbf{n \geq 0},  \\
	& \, \,  \hat{\sigma} \mathbf{y} \geq \mathbf{0}, \, \, \text{with}  \, \,  \mathbf{R}\cdot\mathbf{y} < 0.
	\label{eq:y} 
	\end{align}
	System \eqref{eq:n} consists of $N$ equations and $\alpha N$ inequalities while system \eqref{eq:y} consists of $N\alpha+1$ inequalities. With each solution $\mathbf{y}$ to \eqref{eq:y} also $\lambda \mathbf{y}$ with positive $\lambda$ is a solution. In order to eliminate this trivial degeneracy it is convenient to impose the spherical constraint $\|\mathbf{y}\|^2=N$ on the solution vectors $\mathbf{y}$. To characterize the solution space of \eqref{eq:y} we introduce the fractional volume of normalized vectors $\mathbf{y}$ solving this system:
	\begin{align}
	\Omega (\hat{\sigma},\mathbf{R}):=
	\frac{\int^\infty_{-\infty} \prod_i dy_i\, \delta \left(\sum_i y_i^2-N\right)
		\Theta \left(-\frac{1}{\sqrt{N}}\sum_i  R_i y_i \right)
		\prod_\mu\Theta \left( \frac{1}{\sqrt{N}} \sum_i \sigma_{\mu i} y_i \right)  }{\int^\infty_{-\infty} \prod_i dy_i \,\delta(\sum_i y_i^2-N)}.
	\label{Equ: Volume of solutions}
	\end{align}
	If this volume is zero the linear equation system (\ref{Equ: Linear Equation System}) possesses a non-negative solution, if it is positive there is no such solution to \eqref{Equ: Linear Equation System}.
	
	\section{The typical fractional volume}
	The fractional volume $\Omega$ is a random quantity due to its dependence on the random parameters $\hat{\sigma}$ and $\mathbf{R}$. Because of its product structure its logarithm is expected to be self-averaging. Therefore, the central quantity of interest is the averaged intensive entropy
	\begin{equation}\label{defS}
	S(\alpha,p,r^2):=\lim_{N\to\infty}\frac{1}{N}
	\langle\log \Omega(\hat{\sigma},\mathbf{R})\rangle_{\hat{\sigma},\mathbf{R}}.
	\end{equation}
	It may be calculated using the replica trick based on the identity:
	\begin{equation}\label{replica}
	\left< \log  \Omega \right>_{\hat{\sigma},\mathbf{R}}=\lim_{n\rightarrow 0}\frac{ \left<\Omega^n\right>_{\hat{\sigma},\mathbf{R}}-1}{n}.
	\end{equation}
	$\langle\Omega^n\rangle_{\hat{\sigma},\mathbf{R}}$ is determined for $n\in N$ and the result needs to be continued in a meaningful way to real $n$ in order to perform the limit $n\to 0$. For $n\in N$ we have
	\begin{align}
	\Omega^n(\alpha,\hat{\sigma},\mathbf{R})=\int^\infty_{-\infty} \prod_{i,a} \frac{dy^a_i}{\sqrt{2 \pi e}} \prod_a \delta \left(\sum_i (y^a_i)^2-N \right)\prod_{\mu,a} \Theta \left( \frac{1}{\sqrt{N}} \sum_i \sigma_{\mu i} y^a_i \right)\prod_a \Theta \left(-\frac{1}{\sqrt{N}}\sum_i R_i y^a_i \right),
	\label{Equ:replica Volume of solutions}
	\end{align}
	where the replica index $a$ runs from 1 to $n$ and the denominator $\sqrt{2 \pi e}$ is due to the normalization in \eqref{Equ: Volume of solutions}.
	
	Using standard techniques [1] we replace the $\delta$- and $\Theta$-functions by their integral representations:
	\begin{align}
	&\prod_a \delta\left(\sum_i (y^a_i)^2-N \right)=\int \prod_a \frac{dE^a}{4 \pi} 
	\text{exp} \left(\frac{i}{2}\sum_a E^a(\sum_i (y^a_i)^2-N)\right),
	\\
	&\prod_a \Theta \left(-\frac{1}{\sqrt{N}}\sum_i R_i y^a_i \right)=\int^\infty_0\prod_a d\eta^a \int \prod_a \frac{d\hat{\eta}^a}{2 \pi/N }\text{exp}\left( i N  \sum_a \hat{\eta}^a\Big(\eta^a+\frac{1}{\sqrt{N}}\sum_i y^a_i+\frac{1}{\sqrt{N}}\sum_i \delta R_i y^a_i\Big)\right),
	\\
	&\prod_{\mu,a} \Theta \left( \frac{1}{\sqrt{N}} \sum_i \sigma_{\mu i} y^a_i \right)=\prod_\mu \int^\infty_0 \prod_a d\vartheta^a_\mu \int \prod_a \frac{d\hat{\vartheta}^a_\mu}{2 \pi} \text{exp} \left(i\sum_a \hat{\vartheta}^a_\mu\Big(\vartheta^a_\mu-\frac{1}{\sqrt{N}}\sum_i\sigma_{\mu i}y^a_i\Big) \right).
	\end{align}
	The averages over $\hat{\sigma}$ and $\delta R_i$ then yield:
	\begin{align}
	\prod_i \left< \text{exp} \left(-\frac{i}{\sqrt{N}}\sum_{a} \sigma_{\mu i} \hat{\vartheta}^a_\mu y^a_i\right)\right>&=\text{exp}\biggl(-\frac{ip}{\sqrt{N}}\sum_{i,a} \hat{\vartheta}^a_\mu y^a_i-\frac{p(1-p)}{2}\sum_a (\hat{\vartheta}^a_\mu)^2
	\\
	&\qquad\qquad-\frac{p(1-p)}{2N} \sum_{(a,b)} \hat{\vartheta}^a_\mu \hat{\vartheta}^b_\mu \sum_i y^a_i y^b_i +\mathcal{O}(N^{-1/2})  \biggr),
	\\
	\prod_i	\left<\text{exp} \left(i\sqrt{N} \delta R_i\sum_a \hat{\eta}^a y^a_i\right) \right>&=\text{exp}\left(-\frac{N}{2}r^2\sum_a (\hat{\eta}^a)^2-\frac{1}{2}r^2 \sum_{(a,b)} \hat{\eta}^a\hat{\eta}^b \sum_i y^a_i  y^b_i \right).
	\end{align}
	Here $\sum_{(a,b)} ...$ means that the diagonal terms, $a=b$, are excluded from the sum. 
	The integrals over the $y_i$ and those over the auxiliary parameters $\vartheta^a_\mu,\hat{\vartheta}^a_\mu, \eta^a$ and $\hat{\eta}^a$ can be decoupled by introducing the order parameters
	\begin{equation}
	m^a=\frac{1}{\sqrt{N}}\sum_i y^a_i\qquad\mathrm{and}\qquad 
	q^{ab}=\frac{1}{N}\sum_i y^a_i y^b_i\quad\mathrm{for}\; a<b,
	\end{equation} 
	via appropriate $\delta$-functions
	\begin{align}
		\delta(m^a-\frac{1}{\sqrt{N}}\sum_i y^a_i)=\int \frac{ d\hat{m}^a}{2 \pi /\sqrt{N}} \text{exp}\left(i \sqrt{N} \hat{m}^a(m^a-\frac{1}{\sqrt{N}}\sum_i y^a_i) \right),
	\\
	\delta(q^{ab}-\frac{1}{N}\sum_i y^a_i y^b_i)=\int \frac{ d\hat{q}^{ab}}{2 \pi /N} \text{exp}\left(i N \hat{q}^{ab}(q^{ab}-\frac{1}{N}\sum_i y^a_i y^b_i) \right).
	\end{align}
	
	The expression for the $n$-th power of the fractional volume then acquires the form
	\begin{align}\nonumber
	\left< \Omega^n \right>=\int \prod_a &\frac{dm^a d\hat{m}^a}{2 \pi /\sqrt{N}}
	\int \prod_{a<b} \frac{dq^{ab} d\hat{q}^{ab}}{2 \pi /N}\int \prod_a \frac{dE^a}{4 \pi} 
	\int^\infty_0\prod_a d\eta^a \int \prod_a \frac{d\hat{\eta}^a}{2 \pi/N}\\\nonumber
	& \quad\text{exp} 
	\Big( N\Big[\frac{i}{\sqrt{N}}\sum_a m^a\Hat{m}^a+i\sum_{a<b}q^{ab}\hat{q}^{ab}-\frac{i}{2}\sum_a E^a 
	+i \sum_a \hat{\eta}^a \eta^a+ i  \sum \hat{\eta}^a m^a\\
	& \quad -\frac{r^2}{2}\sum_a (\hat{\eta}^a)^2-\frac{r^2}{2} \sum_{(a,b)} \hat{\eta}^a\hat{\eta}^b q^{ab} +\alpha G_E(m^a,q^{ab})+G_S(E^a,\hat{m}^a,\hat{q}^{ab})\Big]\Big)\label{Omn},
	\end{align}
	with the auxiliary functions
	\begin{equation}\label{defGE}
	G_E(m^a,q^{ab})=\log \int^\infty_0 \prod_a d\vartheta^a \int \prod_a \frac{d\hat{\vartheta}^a}{2 \pi}\, \text{exp} \Big(i\sum_a \hat{\vartheta}^a \vartheta^a-ip \sum_a \hat{\vartheta}^a m^a -\frac{p(1-p)}{2}\sum_a (\hat{\vartheta}^a)^2-\frac{p(1-p)}{2} \sum_{(a,b)} \hat{\vartheta}^a \hat{\vartheta}^b q^{ab} \Big),
	\end{equation} 
	and 
	\begin{equation}\label{defGS}
	G_S(E^a,\hat{m}^a,\hat{q}^{ab})=\log \int \prod_a \frac{dy^a}{\sqrt{2 \pi e}}\, \text{exp} 
	\Big(\frac{i}{2}\sum_aE^a (y^a)^2-i\sum_a \Hat{m}^ay^a-i\sum_{a<b}\Hat{q}^{ab}y^a y^b \Big).
	\end{equation} 
	To determine the entropy \eqref{defS} we only need the asymptotics of this expression for $N\to\infty$ so that the integrals in \eqref{Omn} may be calculated by the saddle-point method. The term $\frac{1}{\sqrt{N}}\sum_a m^a \hat{m}^a$ can be neglected in this limit and will be dropped.
	
	We assume a replica-symmetric saddle-point and make the ans\"atze
	\begin{align}
	m^a=& \, m, \quad  i\hat{m}^a=-\hat{m}, \quad iE^a=-E,\quad 
	\eta^a=\eta,\quad i\hat{\eta}^a=\hat{\eta} &\forall a,
	\\
	q^{ab}=& \, q, \, \, \, \, \, \,  i\hat{q}^{ab}=-\hat{q} &\forall a\neq b.
	\end{align}
	
	Using this replica-symmetric structure at the saddle-point and keeping in mind that for the final limit $n\to 0$  only terms up to order $n$ are needed the expressions for $G_E$ and $G_S$ may be simplified. We find
	\begin{align}
	G_E(m,q)&=\log \int^\infty_0 \prod_a d\vartheta^a \int \prod_a \frac{d\hat{\vartheta}^a}{2 \pi} \text{exp} 
	\Big(i\sum_a\hat{\vartheta}^a \vartheta^a-ipm \sum_a \hat{\vartheta}^a 
	-\frac{p(1-p)}{2} \big[(1-q)\sum_a (\hat{\vartheta}^a)^2+q (\sum_a \hat{\vartheta}^a)^2 \big]\Big)\\
	&=\log \int Dt \left[\int^\infty_0 d\vartheta \int \frac{d\hat{\vartheta}}{2 \pi} \text{exp} 
	\Big(i\hat{\vartheta} \vartheta-ipm \hat{\vartheta} -\frac{p(1-p)}{2}(1-q)\hat{\vartheta}^2
	+i\,t\,\sqrt{p(1-p)q}\, \hat{\vartheta}\Big)\right]^n\\
	&=\log\int Dt \left[\int^\infty_0 \frac{d\vartheta}{\sqrt{2\pi p(1-p)(1-q)}}\, \text{exp} 
	\Big(-\frac{\big(\vartheta-pm +t\sqrt{p(1-p)q}\,\big)^2}{2 p(1-p)(1-q)}\Big)\right]^n\\
	&=\log\int Dt\, H^n\Big(\sqrt{\frac{q}{1-q}}\,t-\frac{pm}{\sqrt{p(1-p)(1-q)}}\Big)
	=\log\Big(1+n\int Dt\, \log H \Big(\frac{\sqrt{q}\,t-\kappa}{\sqrt{1-q}}\Big) +O(n^2)\Big)\\
	&=n \int Dt\, \log H \Big(\frac{\sqrt{q}\,t-\kappa}{\sqrt{1-q}}\Big) +O(n^2).
	\end{align}
	Here the Gaussian measure $Dt:=dt/\sqrt{2\pi}\,e^{-t^2/2}$ was introduced to perform the Hubbard-Stratonovich transform in the second line and the abbreviations 
	\begin{equation}
	H(x):=\int_x^\infty Dt\qquad\mathrm{and}\qquad\kappa:=m\sqrt{\frac{p}{1-p}}
	\end{equation} 
	were used. Similar manipulations yield for $G_S$
	\begin{align}
	G_S(E,\hat{m},\hat{q})&=\log \int \prod_a \frac{dy^a}{\sqrt{2 \pi e}}\, \text{exp} 
	\Big(-\frac{E}{2} \sum_a (y^a)^2+\Hat{m}\sum_a y^a+ \frac{\hat{q}}{2}\sum_{(a,b)} y^a y^b \Big)\\
	&=\log \int Dz \left[\frac{1}{\sqrt{e(E+\hat{q})}} \, 
	\text{exp}\Big(\frac{(\hat{m}+\sqrt{\hat{q}}z)^2}{2(E+\hat{q})}\Big)\right]^n\\
	&=-\frac{n}{2}\Big(1+\log(E+\hat{q})\Big) + \frac{n}{2}\frac{\hat{m}^2+\hat{q}}{E+\hat{q}}+O(n^2).
	\end{align}
	
	Simplifying also the remaining terms in \eqref{Omn} for a replica-symmetric saddle-point and using \eqref{defS} and \eqref{replica} we finally get
	\begin{align}\nonumber
	S(\alpha,p,r^2)
	&=\mathrm{extr}\left[\frac{q\hat{q}}{2}+\frac{E}{2}+\eta\hat{\eta}+\hat{\eta}\kappa\sqrt{\frac{1-p}{p}}
	+\frac{r^2}{2}(1-q)\hat{\eta}^2-\frac{1}{2}
	-\frac{1}{2}\log(E+\hat{q})+\frac{\hat{m}^2+\hat{q}}{2(E+\hat{q})}\right.\\ 
	&\qquad\qquad\qquad\left.+\alpha\int Dt\, \log H \Big(\frac{\sqrt{q}\,t-\kappa}{\sqrt{1-q}}\Big)\right]
	\label{h1},
	\end{align}
	where the extremum is over $E,\kappa,\hat{m},q,\hat{q},\eta$ and $\hat{\eta}$. 
	
	Except for $\kappa$ and $q$ the saddle-point equations are algebraic and may be used to eliminate the respective variables. From the saddle point equation for $\hat{m}$ we immediately get $\hat{m}=0$, those with respect to $E$ and $\hat{q}$ give 
	\begin{equation}
	E=\frac{1-2q}{(1-q)^2} \qquad\mathrm{and}\qquad \hat{q}=\frac{q}{(1-q)^2}.
	\end{equation} 
	Plugging these expressions in \eqref{h1} yields 
	\begin{equation}\label{h2}
	S(\alpha,p,r^2)
	=\mathrm{extr}\left[\frac{1}{2}\log(1-q)+\frac{q}{2(1-q)}+\eta\hat{\eta}+\hat{\eta}\kappa\sqrt{\frac{1-p}{p}}
	+\frac{r^2}{2}(1-q)\hat{\eta}^2 
	+\alpha\int Dt\, \log H \Big(\frac{\sqrt{q}\,t-\kappa}{\sqrt{1-q}}\Big)\right].
	\end{equation}
	The extremum in $\eta$ is somewhat unusual, it is at the lower boundary of the integration region, i.e. at $\eta=0^+$. Intuitively this means that we satisfy the inequality $\mathbf{R}\cdot\mathbf{y}<0$ in \eqref{eq:y} by $\mathbf{R}\cdot\mathbf{y}=0^-$. It is not unreasonable that the solution volume is maximized for large $N$ when the inequality is pushed to its limit. Setting the derivative with respect to $\hat{\eta}$ to zero gives
	\begin{equation}
	\hat{\eta}=-\frac{\kappa}{r^2(1-q)}\sqrt{\frac{1-p}{p}}
	\end{equation} 
	and we therefore finally find
	\begin{equation}\label{Sres}
	S(\alpha,p,r^2)=\mathrm{extr}\left[\frac{1}{2}\log(1-q)+\frac{q}{2(1-q)}-\frac{\kappa^2(1-p)}{2pr^2(1-q)}
	+\alpha\int Dt\, \log H \Big(\frac{\sqrt{q}\,t-\kappa}{\sqrt{1-q}}\Big)\right],
	\end{equation} 
	where the extremum remains to be taken only over $\kappa$ and $q$. This is the expression given in the main text. 
	
	\section{The phase transition}
	At the transition the solution volume $\Omega$ shrinks to zero and the overlap $q$ between different solutions $\mathbf{y}$ has to tend to one. Then to leading order in $1/(1-q)$ we have 
	\begin{equation}
	\log H \Big(\frac{\sqrt{q}\,t-\kappa}{\sqrt{1-q}}\Big)\sim
	\begin{cases}
	-\frac{(t-\kappa)^2}{2(1-q)} & \text{if}\quad t>\kappa,\\
	0              & \text{otherwise.}
	\end{cases}
	\end{equation} 
	and therefore
	\begin{equation}
	\int Dt\, \log H \Big(\frac{\sqrt{q}\,t-\kappa}{\sqrt{1-q}}\Big)
	\sim -\frac{1}{2(1-q)}\int_\kappa^\infty Dt\,(t-\kappa)^2=:-\frac{1}{2(1-q)}\, I(\kappa).
	\end{equation} 
	Keeping only the most divergent terms hence gives near the transition 
	\begin{equation}
	S(\alpha,p,r^2)\sim\mathrm{extr}\left[\frac{1}{2(1-q)}-\frac{(1-p)}{2 \, p \, r^2}\frac{\kappa^2}{(1-q)}-\frac{\alpha}{2(1-q)} I(\kappa)\right].
	\end{equation} 
	The saddle-point equation with respect to $q$ gives
	\begin{equation}\label{h4}
	1-\frac{(1-p)}{p \, r^2} \kappa^2=\alpha I(\kappa)
	\end{equation} 
	resulting in 
	\begin{equation}\label{res1}
	r^2= \frac{1-p}{p}\frac{\kappa^2}{1-\alpha I(\kappa)}.
	\end{equation}
	The saddle-point equation with respect to $\kappa$ gives 
	\begin{equation}
	-\frac{(1-p)}{p \, r^2}\kappa=\frac{\alpha}{2} \frac{d I}{d\kappa}
	=\frac{\alpha}{\kappa}\big(I(\kappa)-H(\kappa)\big),
	\end{equation} 
	which together with \eqref{h4} yields
	\begin{equation}\label{res2}
	\alpha H(\kappa)=1.
	\end{equation} 
	Eqs.~\eqref{res1} and \eqref{res2} give a parametric description of the transition line in the $\alpha$-$r^2$-plane. They are identical to the equations found in [2] for the critical line.

	\section{The competitive exclusion principle and the limit $\epsilon \rightarrow 0^+$}
	We are interested in the stationary states of the model given by
	\begin{equation}
	\frac{dn_{\mu}}{dt}=n_{\mu} (\boldsymbol{\sigma}_\mu \cdot \boldsymbol{h}-\chi_\mu)=0 \, \, \, \text{for all } \mu=1\dots S. 
	\label{Eq.: Steady states}
	\end{equation}
	
	This equation implies that in the stationary states for each component $\mu$ either the population has to vanish $n_\mu=0$ or the equation $\boldsymbol{\sigma}_\mu \cdot \boldsymbol{h}=\chi_\mu$ has to be fulfilled. For a general vector of thresholds $\boldsymbol{\chi}$ and a full-rank matrix of metabolic strategies $\hat{\sigma}$, the overdetermined linear equation system
	\begin{equation}
	\boldsymbol{\hat{\sigma}} \cdot \boldsymbol{h}=\boldsymbol{\chi}, \, \, \boldsymbol{\hat{\sigma}} \in \mathbb{R}^{N\alpha \times N}, \, \boldsymbol{\chi} \in \mathbb{R}^{N\alpha},
	\label{Eq: Overdetermined Equation System}
	\end{equation}
	will have no solution $\boldsymbol{h}$. This means that not all species can survive at the same time. For large system sizes a random $\boldsymbol{\hat{\sigma}}$ has full rank  with overwhelming probability. It is then always possible to find a vector of resource availabilities $\boldsymbol{h}$ such that $N$ equations are fulfilled. Therefore, the competitive exclusion principle is fulfilled.
	\newline
	\\
	In the model the following choice of thresholds is made
	\begin{equation}
	\chi_\mu=\sum_i \sigma_{\mu i} + \epsilon x_\mu,
	\end{equation}
	where $x_\mu$ is a Gaussian random variable with zero mean and variance 
	one and $\epsilon$ is a small positive number quantifying the scatter of the thresholds. With this choice, Eq.~(\ref{Eq: Overdetermined Equation System}) takes the form:
	\begin{equation}
	\boldsymbol{\hat{\sigma}} \cdot (\boldsymbol{{h}}-\boldsymbol{1})=\epsilon \boldsymbol{x}.
	\end{equation}
	
	For $\epsilon=0$ the vector $\boldsymbol{\chi}$ is a linear combination of the columns of $\boldsymbol{\hat{\sigma}}$ and one observes a highly degenerated situation. Then, ${h}_i=1$ for all $i$ is a solution of the system which makes the bracket in Eq.~(\ref{Eq.: Steady states}) zero for all $\mu$ and hence all population sizes can be non-zero in the stationary state. 
	On the contrary, if $\epsilon >0 $ the vector $\boldsymbol{{\chi}}$ is typically linearly independent of $\boldsymbol{\hat{\sigma}}$ and again only $N$ equations of the system can be fulfilled. Since we are not interested in the degenerated states and the phase transition only occurs for vanishing scatter we consider the case $\epsilon \rightarrow 0^+$. For finite $\epsilon$ the phase transition is replaced by a crossover, see [2].
	\\
	
	Assume that the system is in the S-phase and $N$ species survive in the stationary state. Then the equation
	\begin{equation}
	\boldsymbol{\hat{\sigma}_S} \cdot (\boldsymbol{{h}}-\boldsymbol{1})=\epsilon \boldsymbol{x_S},
	\label{Eq: Survivor equation}
	\end{equation}
	with the matrix of the surviving metabolic strategies $\hat{\sigma}_S \in \mathbb{R}^{N \times N}$ and the scatter of thresholds of the surviving species $\epsilon \, \mathbf{x}_S \in \mathbb{R}^N$ is fulfilled. For large system sizes the matrix $\hat{\sigma}_S$ will have full rank with overwhelming probability such that it always has an inverse and we can write
	\begin{equation}
	(\boldsymbol{{h}}-\boldsymbol{1})=\epsilon \boldsymbol{\hat{\sigma}_S}^{-1}{\boldsymbol{x}_S}.
	\label{Eq: Solution for N equations}
	\end{equation}
	
	In the limit of vanishing scattering $\epsilon \rightarrow 0^+$ the components of $h_i$ equal one. This shows that the survival of $N$ species implies that all resource availabilities are equal to one. 
	
	Now consider the case that the system is in the V-phase and $N_S<N$ species survive. Then, $\sigma_S\in \mathbb{R}^{N_S \times N}$ and $\epsilon \, \mathbf{x}_S \in \mathbb{R}^{N_S}$. This means that Eq.~(\ref{Eq: Survivor equation}) becomes an underdetermined equation system which does not have a unique solution and the components $h_i$ do not necessarily approach one for $\epsilon \rightarrow 0^+$.

	\section{Finite size analysis}
	The data in Figure 3 was obtained in the following way: For each choice of $\alpha$ and $N$ we created  400 random realizations of the linear equation system $$\hat{\sigma}^T \mathbf{n}=\mathbf{R}.$$ Using the non-negative least squares solver nnls of the scipy.optimize package in Python it was checked whether the system possesses a non-negative solution $\mathbf{n}$. The data points show the average number of cases in which such a solution was found.
	\\
	To fit the data for the respective system sizes $N$ we used the hyperbolic tangent function
	
	\begin{equation}
	\text{Fraction of realizations with solution}(\alpha)=\frac{1}{2}+\frac{1}{2} \tanh(a \alpha+b), \nonumber 
	\end{equation}
	where the parameters $a$ and $b$ were fitted. 
	\\
	The finite-size approximation of the critical $\alpha_c$ is given by the value of $\alpha$ where the hyperbolic tangent vanishes and half of the systems posses a solution, i.e.  $\alpha_c(N)=-b/a$. The fits in the insets were done using a power law
	$$
	\alpha_c (N)=a'+b' \, N^{c},
	$$
	where $a',b'$ and $c$ are fit parameters.
\vspace{1cm}
\newline
\\
\textbf{References}
\\
{[1]}  Engel A. and Van den Broeck C., Statistical mechanics of learning (Cambridge University Press) 2001.
\\
{[2]}  Tikhonov M. and Monasson R., Phys. Rev. Lett., \textbf{118} (2017) 048103.

\end{document}